
%
%
\input harvmac
\def\sc{\scriptstyle}
\def\scsc{\scriptscriptstyle}
\def\sfrac#1#2{{\sc {#1}\over{#2}}}

\def\nl{\hfill\break}
\def\bar{\overline}
\def\lam{\lambda}
\def\ep{\epsilon}

\def\tht{\theta}

\def\shalf{{\scriptstyle {1\over 2}}}

\def\sN{{\scriptscriptstyle N}}
\def\sE{{\scriptscriptstyle E}}

\def\ZZ{{\bf Z}}
\def\IC{{\ontopss{$\scriptscriptstyle \mid$}{{\rm C}}{8.0}{0.415}}}
\def\ontopss#1#2#3#4{\raise#4ex \hbox{#1}\mkern-#3mu {#2}}
%

\nref\rZamflow{A.B.~Zamolodchikov, Sov.~J.~Nucl.~Phys. 46 (1987) 1090.}
\nref\rLudCar{A.~Ludwig and J.L.~Cardy, Nucl.~Phys.~B285 (1987) 687.}
\nref\rFQSGKO{D.~Friedan, Z.~Qiu and S.H.~Shenker, Phys.~Rev.~Lett.~52 (1984)
 1575; \nl  P.~Goddard, A.~Kent and D.~Olive, Phys.~Lett.~152B (1985) 88.}
\nref\rBPZ{A.A.~Belavin, A.M.~Polyakov and A.B.~Zamolodchikov,
  Nucl.~Phys.~B241 (1984) 333.}
\nref\rCarMI{J.L.~Cardy, Nucl.~Phys.~B270 (1986) 186;
  {\it ibid.}~B275 (1986) 200.}
\nref\rCIZ{A.~Cappelli, C.~Itzykson and J.-B.~Zuber, Nucl.~Phys.~B280 (1987)
  445.}
\nref\rTanig{A.B.~Zamolodchikov, Adv.~Stud.~Pure Math.~19 (1989) 1.}
\nref\rkink{A.B.~Zamolodchikov, Landau Institute preprint (1989);\nl
  C.~Ahn, D.~Bernard and A.~LeClair, Nucl.~Phys.~B346 (1990) 409  and
  references therein;
  T.~Eguchi and S.-K.~Yang, Phys.~Lett.~235B (1990) 282;\nl
  N.Yu.~Reshetikhin and F.A.~Smirnov, Comm.~Math.~Phys.~131 (1990) 157.}
\nref\rHuse{D.A.~Huse, Phys.~Rev.~B30 (1984) 3908.}
\nref\rZamLG{A.B.~Zamolodchikov, Sov.~J.~Nucl.~Phys.~46 (1987) 1090.}
\nref\rKMS{D.~A.~Kastor, E.~J.~Martinec and S.~H.~Shenker,
 Nucl.~Phys.~B316 (1989) 590.}
\nref\rNeqtwo{E.J.~Martinec, Phys.~Lett.~217B (1989) 431;\nl  C.~Vafa and
 N.P.~Warner, Phys.~Lett.~218B (1989) 51.}
\nref\rCapLat{A.~Cappelli and J.I.~Latorre, Nucl.~Phys.~B340 (1990) 659.}
\nref\rKLi{K.~Li, Phys.~Lett.~219B (1989) 297.}
\nref\rZams{A.B.~Zamolodchikov and Al.B.~Zamolodchikov, Ann.~Phys.~120
(1980) 253.}
\nref\rourvi{T.R.~Klassen and E.~Melzer, ``Finite-size Effects as a Probe
 of Non-perturbative Physics'', Chicago/Miami preprint EFI 91-25/UMTG-164
 (1991), to appear in the proceedings of the Miami NATO workshop
{\it Quantum Field Theory, Statistical Mechanics, Quantum Groups, and Topology}
 (January 1991), T.~Curtright {\it et al}~(eds.), Plenum.}
\nref\rZamtba{Al.B.~Zamolodchikov, Nucl.~Phys.~B342 (1990) 695.}
\nref\rourii{T.R.~Klassen and E.~Melzer, Nucl.~Phys.~B338 (1990) 485.}
\nref\rouriii{T.R.~Klassen and E.~Melzer, Nucl.~Phys.~B350 (1990) 635.}
\nref\rRSOStba{Al.B.~Zamolodchikov, Nucl.~Phys.~B358 (1991) 497.}
\nref\rmlesstba{Al.B.~Zamolodchikov, Nucl.~Phys.~B358 (1991) 524.}
\nref\rourv{T.R.~Klassen and E.~Melzer, ``Spectral Flow between
  Conformal Field Theories in 1+1 Dimensions'', Chicago/Miami preprint
  EFI 91-17/UMTG-162 (1991), Nucl.~Phys.~B (in  press).}
\nref\rZNtba{V.A.~Fateev and Al.B.~Zamolodchikov, ``Integrable Perturbations
 of $\ZZ_\sN$ Parafermion Models and $O(3)$ Sigma Model'',
 Ecole Normale preprint ENS-LPS-341 (1991).}
\nref\rZFparaf{A.B.~Zamolodchikov and V.A.~Fateev, Sov.~Phys.~JETP 62 (1985)
  215;\nl JETP 63 (1986) 913.}
\nref\reMart{M.J.~Martins, Phys.~Lett.~257B (1991) 317.}
\nref\rMartcomp{M.J.~Martins, Phys.~Rev.~Lett.~67 (1991) 419.}
\nref\rFend{P.~Fendley, ``Excited-state Thermodynamics'', Boston preprint
  BUHEP-91-16 (1991).}
\nref\rorbif{P.~Fendley and P.~Ginsparg, Nucl.~Phys.~B324 (1989) 549.}
\nref\rfssc{H.W.J.~Bl\"ote, J.L.~Cardy and M.P.~Nightingale,
 Phys.~Rev.~Lett.~56 (1986) 742;\nl
 I.~Affleck, Phys.~Rev.~Lett.~56 (1986) 746.}
\nref\rremarks{Al.B.~Zamolodchikov, Phys.~Lett.~253B (1991) 391.}
\nref\rGep{D.~Gepner, Nucl.~Phys.~B287 (1987) 111.}
\nref\rPetk{V.B.~Petkova, Int.~J.~Mod.~Phys.~A3 (1988) 2945;
  Phys.~Lett.~225B (1989) 357.}
\nref\rFuKl{J.~Fuchs and A.~Klemm, Ann.~Phys.~194 (1989) 303;\nl
  J.~Fuchs, private communication.}
\nref\rDotFatiii{Vl.S.~Dotsenko and V.A.~Fateev, Phys.~Lett.~154B (1985) 291.}
\nref\rstairs{Al.B.~Zamolodchikov, ``Resonance Factorized Scattering and
 Roaming Trajectories'', Ecole Normale preprint ENS-LPS-355 (1991).}
\nref\rYZ{V.P.~Yurov and Al.B.~Zamolodchikov, Int.~J.~Mod.~Phys.~A5 (1990)
 3221.}
\nref\rDotsCPT{Vl.S.~Dotsenko, Nucl.~Phys.~B314 (1989) 687.}
\nref\rIFT{T.T.~Wu, B.M.~McCoy, C.A.~Tracy and E.~Baruch, Phys.~Rev.~B13
 (1976) 316.}
\nref\rRav{F.~Ravanini, ``RG Flows of Non-diagonal Minimal Models Perturbed
  by $\phi_{1,3}$'', Paris preprint SPhT/91-147 (October 1991).}
\nref\rGhoSen{D.~Ghoshal and A.~Sen, Phys.~Lett.~265B (1991) 295.}

\Title{\vbox{\baselineskip12pt\hbox{CLNS-91/1111}\hbox{ITP-SB-91-57}  }}
{\vbox{\centerline{RG Flows in the $D$-Series of Minimal CFTs}}}

\centerline{Timothy R.~Klassen\foot{email: klassen@strange.tn.cornell.edu}}
\smallskip\centerline{\it Newman Laboratory for Nuclear Studies}
\smallskip\centerline{\it Cornell University, Ithaca, NY 14853}
\medskip\centerline{and}
\medskip\centerline{Ezer Melzer\foot{email: melzer@max.physics.sunysb.edu}}
\smallskip\centerline{\it Institute for Theoretical Physics}
\smallskip\centerline{\it SUNY, Stony Brook,  NY 11794-3840}
\vskip 9mm

Using results of
the thermodynamic Bethe Ansatz approach and conformal perturbation
theory we argue that the $\phi_{1,3}$-perturbation of a unitary minimal
$(1+1)$-dimensional conformal field theory (CFT)
in the $D$-series of modular invariant partition functions induces
a renormalization group (RG) flow to the next-lower
model in the $D$-series. An exception is the first model in the series, the
3-state Potts CFT, which under the
$\ZZ_2$-even $\phi_{1,3}$-perturbation flows to the tricritical Ising
CFT, the second model in the $A$-series.
We present arguments that in the $A$-series
flow corresponding to this exceptional case, interpolating between
the tetracritical and the tricritical Ising CFT,
the IR fixed point is approached from ``exactly the opposite direction''.
Our results indicate how (most of) the relevant conformal fields
evolve from the UV to the IR CFT.

\Date{October 1991}
\vfill\eject

\newsec{Introduction}
\ftno=0

Among the few aspects of quantum field theories (QFTs) often amenable to
analytical study is their behaviour at large and/or short distances.
The asymptotic behaviour in these regimes is governed by
(possibly trivial) fixed points of the
renormalization group (RG), and RG-improved perturbation theory can provide
important insights into the full non-perturbative behaviour of the theory.

A particularly
interesting situation arises if both the ultraviolet (UV) and
infrared (IR) fixed points of
a QFT are nontrivial.  This is interesting, from a QFT point of view
because RG flows between such fixed points provide  some understanding
of the topology of the space of QFTs, and from a statistical mechanics
viewpoint because it yields  examples of the
cross-over between different universality classes of critical phenomena.
 One would like to understand such RG flows in detail,
{\it e.g.}~not just between which theories the flow
interpolates but also how specific operators evolve from the
UV to the IR.

In 1+1 dimensions (where the subject may have implications
also for the non-perturbative formulation of string theory)
it is possible to study some of these questions explicitly.
The best known examples of RG flows between non-trivial fixed points are
those~\rZamflow\rLudCar\
induced by $\phi_{1,3}$-perturbations of the unitary~\rFQSGKO ~minimal
models~\rBPZ ~of central charge $c_m=1-{6\over m(m+1)}$
{}~with diagonal modular invariant partition functions
(MIPFs)~\rCarMI.
Here we study the $\phi_{1,3}$-induced flows between the unitary minimal models
with {\it non-diagonal} MIPFs.

Denote a unitary minimal model of central charge $c_m$ and given MIPF
in the $ADE$ classification~\rCIZ ~by $X_m$,
with $m=3,4,5,\ldots$ for $X=A$, ~$m=5,6,7,\ldots$ for $X=D$, and
{}~$m=11,12,17,18,29,30$ for $X=E$.
Recall that unlike the $A$-models, the $D$- and $E$-models
have non-diagonal MIPFs.
Also, except for the $E$-models with $m=11,17,29$ all
the models contain $\phi_{1,3}$, which is the least relevant
(spinless, primary) field.\foot{$D_5$, the 3-state Potts CFT~\rCarMI,
is special in that its spectrum contains two copies of $\phi_{1,3}$ that
can be taken to be even ($\phi_{1,3}^+$)  and odd ($\phi_{1,3}^-$)
with respect to the global $\ZZ_2$ symmetry
of the model; by ``the $\phi_{1,3}$ perturbation'' in this case we will
always mean the $\ZZ_2$-even perturbation. \nl}
Hence, except for the latter three models, one can define the
perturbed CFTs $X_m^{(\pm)}$ via the (euclidean) action
\eqn\pCFT{
  A_{X_m^{(\pm)}} ~=~ A_{X_m^{ }} + \lambda\int d^2 x ~\phi_{1,3}(x)~,}
where the superscripts $(\pm)$ refer to the sign of $\lambda$.
The theories $X_m^{(\pm)}$ are integrable~\rTanig\
non-scale-invariant theories,
the mass scale being proportional to $\lambda^{1/y_m}$ where
{}~$y_m={4\over m+1}$~ is the RG eigenvalue of the field $\phi_{1,3}$
(its scaling dimension is $d_{1,3}^{(m)}=2-y_m$).
Since under repeated fusions  $\phi_{1,3}$ closes on  itself, modulo
irrelevant operators and the identity, renomalization of the UV divergences
in the perturbation theory based on \pCFT\ will not generate additional
counterterms (the counterterm corresponding to the identity does, however,
affect the perturbative expansions we will be interested in --- see sect.~2).

Perturbations in opposite directions of the same
CFT may lead to completely different non-scale-invariant QFTs. $A_m^{(-)}$,
for example, is believed to be a theory of massive kinks~\rkink. On the other
hand,
$A_m^{(+)}$ ~($m\ge 4$) is believed to be massless and
flow to $A_{m-1}$ in the IR, as strongly suggested~\rZamflow\rLudCar ~by
the LG approach together with perturbative RG analysis (the latter applicable
when $m\gg 1$).\foot{
The existence of these flows was first alluded to by Huse~\rHuse, who noticed
a cross-over between different critical behaviours in regime IV of
the integrable RSOS lattice models underlying the QFTs $A_m^{(+)}$. \nl}
We will present   concrete evidence that the
theories $D_m^{(+)}$ describe the flows
\eqn\Dflow{ D_m ~\to ~\cases{A_4     & ~if ~~$m=5$\cr
                             D_{m-1} & ~if ~~$m\ge 6$\cr}~,}
the UV (IR) limits corresponding to $\lambda=0~(\infty)$, respectively.
It is also natural to conjecture that $E_m^{(+)}$, $m=12,18,30$,
describe the flows $E_m \to E_{m-1}$
(see sect.~3.4).

The two techniques traditionally used to study these and other (conjectured)
RG flows
are Landau-Ginzburg (LG) analysis~\rZamLG\rLudCar\rKMS\rNeqtwo\
and  RG-improved perturbative calculations~\rZamflow\rLudCar\rKMS\rCapLat.
Although LG descriptions involving two scalar (real) fields have been
proposed~\rCarMI\rKLi\ for $D_m$ and $E_m$, they do not seem to be useful
when analyzing $\phi_{1,3}$-induced RG flows.
This is not surprising since the LG models involve two {\it strongly
interacting} fields, even for large $m$.
On the other hand,
RG-improved perturbative calculations~\rZamflow\rLudCar\ can
be performed for any $X_m^{(+)}$ provided $m\gg 1$
(which excludes the $E$-models from consideration),
so that the IR fixed point is close to the UV one.
This will be discussed in sect.~3.3.

We will also provide evidence for \Dflow\ --- in particular for small $m$ ---
along quite different lines. This is possible due to
 some recent developments in
the study of the finite-volume spectrum of non-scale-invariant integrable
QFTs in $1+1$ dimensions. The thermodynamic Bethe Ansatz (TBA) technique
allows one to obtain non-linear integral equations for the exact
ground state energy $E_0(R)$ of such a QFT in finite volume $R$, given
its factorizable~\rZams ~$S$-matrix.
The small $R$ behaviour of $E_0(R)$ contains information about the UV CFT,
which can be compared with predictions of conformal perturbation theory (CPT).
This approach has been extensively used in the last two years as a means of
checking the purely massive scattering theories conjectured to describe the
on-shell behaviour of certain perturbed CFTs (see~\rourvi\ for a brief review
of all this,~\rZamtba\rourii\rouriii\ for the details in the case of perturbed
CFTs with diagonal $S$-matrices, and~\rRSOStba\ for the first analysis of a
non-diagonal scattering theory).

Recently~\rmlesstba ~Al.~Zamolodchikov proposed an explicit
scattering theory  of {\it massless} particles
for the non-scale-invariant QFT
$A_4^{(+)}$, presumably~\rKMS\
 interpolating between the tricritical and the  critical Ising CFT.
Comparison of the solutions of the TBA equations with CPT around the UV and IR
CFTs leaves little doubt that the conjectured $S$-matrix is correct.
In~\rmlesstba\ it was further conjectured that
$E_0(R)$ in $A_m^{(+)}$ for $m>4$ is given by a certain
generalization (see sect.~2) of the TBA equations         for $A_4^{(+)}$.
It was checked that the conjectured TBA equations give the expected UV and IR
central charges.
Nevertheless, since for $m>4$ the TBA equations have simply been guessed
(there are not even any conjectures for the $S$-matrices of $A_m^{(+)}$ for
$m>4$), further evidence for the correctness of the equations is required.

In \rourv\ we provided such evidence for $m=5,6,7$.\foot{The difficulty in
performing the numerical analysis of the TBA equations, required for the
comparison with CPT, increases rapidly with $m$. \nl}
  Furthermore, we proposed TBA-like equations for certain
finite-volume {\it excitation} energies in $A_m^{(\pm)}$ for
$m$ even. For $A_m^{(\pm)}$ these equations explicitly demonstrate the flow
of the UV conformal fields $\phi_{2,2}$ and $\phi_{{m\over 2},{m\over 2}}$ to
fields with the same Kac indices in $A_{m-1}$,   
in agreement with expectations based on LG and perturbative RG analyses.
These results provide
further strong (and independent) support for the existence of the
flows concerned, and show the usefulness of the study of
finite-volume spectrum in the context of RG flows in general.

In fact, the ``TBA approach'' has led the authors of~\rZNtba\ to conjecture
new RG flows that has not been proposed before, namely between the
$\ZZ_\sN$-parafermion CFTs~\rZFparaf ~($N\ge 3$) and the minimal models
$A_{\sN+1}$.
The conjectures are based on equations for $E_0(R)$ of the perturbed
parafermion models (the perturbation is by a certain field that breaks the
$\ZZ_\sN$ symmetry, leaving only a global $\ZZ_2$ symmetry in the resulting
theory). Although the equations have been basically obtained by guesswork
and it is desirable to come up with further evidence that
they are correct, previous experience suggests to take them seriously. 
In the particular case $N=3$, where the perturbed $\ZZ_3$-parafermion model is
just the $\phi_{1,3}$-perturbed
3-state Potts CFT $D_5^{(+)}$, the equations for $E_0(R)$ are the same
as the ones proposed in \rmlesstba\ for the ground state energy in
$A_5^{(+)}$, corresponding to the  flow $A_5 \to A_4$.

More generally, the {\it ground state} energy in $X_m^{(\pm)}$
does not depend on the MIPF of the unperturbed theory;
this is predicted by UV-CPT (see sect.~3.2).
Assuming the existence of the flows within the $A$-series, it is
therefore clear that  for {\it all} $m\ge 5$ $D_m^{(+)}$ describes
a flow  of $D_m$ to $X_{m-1}$, but with what MIPF?
Our answer is eq.~\Dflow.

In sect.~2 we will describe in more detail known TBA and CPT results, leading
to our conjecture~\Dflow.  In sect.~3 new evidence from UV and IR CPT
as well as RG-improved CPT
 is presented. We also  discuss in some detail the special case of
$D_5^{(+)}$, arguing that from an IR-CPT point of view, {\it i.e.}~considering
it as a $\phi_{3,1}$-perturbation ($+$ higher corrections) of $A_4$, it is just
the {\it sign} of the $\phi_{3,1}$-coupling that distinguishes it from
$A_5^{(+)}$.
In sect.~4 we briefly discuss our results and methods, concluding with
an outlook on some open questions.
Some observations regarding the appearance of logarithmic terms in CPT
expansions are described in the appendix.

\vskip 4mm

\newsec{TBA and CPT results}

We now describe in more detail the results of the TBA approach
for the theories under consideration.
Define the {\it scaling function} $e(r)=(2\pi)^{-1}RE(\lambda,R)$
corresponding to a generic excitation energy $E(\lambda,R)$ in a one-parameter
($\lambda$, of a specific sign) family of perturbed CFTs.
Here $R$ is the circumference of the cylinder on which the theory is defined
(periodic boundary conditions are assumed), and
$r=MR$ is
a dimensionless scaling parameter, $M$ being the mass scale of the theory.
$M$ is usually chosen in a way that is natural from the point of view of the
off-critical theory, {\it e.g.}~the infinite-volume mass of the lightest
particle if the the theory is purely massive.
It turns out that such a choice is also convenient
for the TBA equations below.  $M$
is related to $\lambda$
through
\eqn\kappadef{ |\lambda| ~=~ \kappa~ M^y~,}
where $\kappa>0$ is a numerical constant
($y$ is the RG eigenvalue of the perturbing field).
$\kappa$
has been determined in many perturbed CFTs numerically, and in a few cases also
analytically  (cf.~the appendix).

At present, all
the TBA-like equations known or conjectured to give the exact scaling functions
of certain excitations in certain integrable theories are of the
``universal'' form
\eqn\eofr{ e(r) ~=~ -{r\over 4\pi^2} \sum_{a=1}^\sN \int_{-\infty}^\infty
   d\theta ~\nu_a(\theta) ~\ln\bigl(1+t_a e^{-\epsilon_a(\theta)}\bigr)~,}
where the $r$-dependent functions $\ep_a(\tht)$ satisfy the equations
\eqn\epseq{ \ep_a(\tht) ~=~ r\nu_a(\tht)-\sum_{b=1}^\sN \int_{-\infty}^\infty
  {d\tht'\over 2\pi} ~K_{ab}(\tht-\tht')
       ~\ln\bigl(1+t_b e^{-\ep_b(\tht')}\bigr)~.}
Here the $t_a$, $a=1,\ldots,N$ (that will be collectively referred to
as the ``type'' $t$ of the TBA system) are certain roots of unity,
the kernel $K$ is a symmetic matrix whose elements are even functions of
$\tht$, exponentially decaying as $|\tht|\to\infty$, and the $\nu_a(\tht)$
are of the form
\eqn\nus{ \nu_a(\tht) \in \left\{\hat{m}_a \cosh\tht,
       ~\shalf\hat{m}_a e^{\pm\tht} \right\} }
where the $\hat{m}_a$ are some non-negative dimensionless parameters
(if non-zero, they are mass ratios in the theory).

Ground state scaling functions $e_0(r)$ always correspond to the ``trivial''
type choice $t=(1,\ldots,1)$. That other choices of type may yield
excitations $e(r)$ in certain theories was first proposed in
\reMart\rourv\rMartcomp.  Fendley~\rFend\ then showed that
when the theory has some global discrete symmetry and
is described by a diagonal  $S$-matrix,
the $e(r)$
corresponding to nontrivial types are the ground state energies of
some sectors of the theory with ``twisted'' boundary conditions.
These $e(r)$ coincide with scaling functions of
excited states in the theory (or some ``orbifolded''~\rorbif\
version of it) with periodic boundary conditions.
With the notable exception of the massless theory $A_4^{(+)}$,
the procedure works only for the spontaneously broken symmetry phase of the
theory, and the excited states obtained become degenerate with the
ground state in infinite volume.  It is not clear
at present how the approach of~\rFend\
can be extended to justify the applicability of the ``change of
type'' prescription in theories with non-diagonal $S$-matrices.

We will not elaborate here on all the specific choices of $N$, $t_a$,
$\nu_a(\tht)$, and $K_{ab}(\tht)$ in \eofr --\epseq ~that are known, or
conjectured, to give rise to certain scaling functions in perturbed CFTs.
Rather, we will specialize  to the cases
we are interested in here, namely the theories $X_m^{(\pm)}$.
Let $e(X_m^{(\pm)},(p,q)|r)$ denote the scaling function of
the energy eigenstate in $X_m^{(\pm)}$ whose UV limit is created by the
spinless primary field $\phi_{p,q}$ in the CFT $X_m$. This
means~\rfssc\rCarMI,
in particular, that $e(X_m^{(\pm)},(p,q)|0)= d_{p,q}^{(m)}-{c_m\over 12}$~
where ~$d_{p,q}^{(m)}={(p(m+1)-qm)^2-1\over 2m(m+1)}$ is the scaling dimension
of $\phi_{p,q}$ and $c_m$ is the central charge of $X_m$.
Recall~\rCIZ ~that in the
CFT $A_m$ all the primary fields are spinless and each $\phi_{p,q}$ with
$1\le q \le p \le m-1$ appears exactly once.\foot{
Alternatively, we can allow all pairs $(p,q)$ with $1\le p\le m-1$ and
$1\le q\le m$, identifying $(p,q)\equiv (m-p,m+1-q)$. \nl}
 The $D_m$ and $E_m$ models do
not contain all of the above spinless primary fields.
Instead, some primary fields
$\phi_{p,q;\bar{p},\bar{q}}$ (with {\it different} left
and right Kac label pairs) with nonzero spin appear, and
some spinless fields are
doubled (the precise field content of the $D_m$ models will be given
in sect.~3.1).
The ground state always corresponds to $(p,q)=(1,1)\equiv (m-1,m)$.

The equations for the ground state scaling function $e(A_m^{(-)},(1,1)|r)$
proposed in~\rRSOStba ~are given by \eofr --\epseq\ with $N=m-2$, all $t_a=1$,
$\nu_1(\tht)=\cosh\tht$, $\nu_{a>1}(\tht)\equiv 0$, and
$K(\tht)=I^{(m-2)}/\cosh\tht$ where $I^{(\sN)}$ is the incidence matrix of the
simple Lie algebra $A_\sN$, {\it i.e.}~$I_{ab}^{(\sN)}=\delta_{a,b-1}+
\delta_{a,b+1}$ for $a,b=1,\ldots,N$. The equations for
$e(A_m^{(+)},(1,1)|r)$ differ from the above only by the choice of
$\nu_1(\tht)$ and $\nu_{m-2}(\tht)$, which are taken to be~\rmlesstba
{}~$\shalf e^\tht$ and $\shalf e^{-\tht}$, respectively. In \rourv\ we then
conjectured that for $m$ even the equations for $e(A_m^{(\pm)},(2,2)|r)$ and
$e(A_m^{(\pm)},({m\over 2},{m\over 2})|r)$  are obtained from those for
$e(A_m^{(\pm)},(1,1)|r)$ simply by  changing the type to
{}~$t=(1,\ldots,1,t_{{m\over 2}-1}=-1,t_{{m\over 2}}=-1,1,\ldots,1)$~ and
{}~$t=(-1,\ldots,-1)$, respectively (the particular case
$e(A_4^{(-)},(2,2)|r)$ of this conjecture appeared also in \rMartcomp ).

In all cases the assignment of the Kac labels to
the solutions of the corresponding TBA-like equations was justified by an
analytic calculation of their $r\to 0$ limits. Further strong support for the
conjecture that the solutions give the correct scaling functions was
provided by numerical calculations and comparison with CPT
in the cases $m=4,5,6,7$~\rourv\  (such analysis of the ground state for $m=4$
was first performed in \rRSOStba\rmlesstba ).
Specifically, CPT based on an action of the form
\pCFT ~predicts the small $r$ expansion of the scaling functions
\eqn\eofrexp{ e(r)~=~e(0)+\sum_{n=1}^\infty a_n r^{ny} +
({\rm term(s)~nonanalytic~in}~r^y, ~{\rm possibly})  ~,}
where $y$ is the RG eigenvalue of the perturbing field and the
{\it CPT coefficients} $a_n$
can be written as
 integrated critical correlators
(see eq.~(3.6) below). In all cases of integrable relevant perturbations of
nontrivial CFTs studied so far, the possible non-analyticity in $r^y$
turns out to be given by a single term.
This term is either proportional to $r^2$, if 2 is not
an integral multiple of $y$, or to $r^2 \ln r$ otherwise.
An $r^2 \ln r$ term arises when integrated correlators
leading to some $a_n$ (namely with $n=2/y$) still diverge after
regularization of their UV divergences
 --- required whenever $y\le 1$ --- by means of analytic continuation in $y$.
Both of these terms can be understood as arising from the RG mixing of
the perturbing field with the identity operator. The precise coefficients
of these terms are determined by a renormalization condition, say
$\lim_{R\to \infty}E_0(R)=0$, which is automatically enforced in the TBA
calculation. No matter what the form of the non-analytic term is,
it is clear that it is the same for all scaling functions in a given model.
It has been determined analytically from the TBA equations in many cases.
(On the other hand, the part of the small $r$ expansion \eofrexp
{}~which is regular in $r^y$ cannot be derived analytically within the TBA
approach by present methods. See~\rremarks, however, for arguments making its
appearance plausible starting from the TBA equations.)

An important consequence of the CPT analysis is
that the scaling functions $e^{(\pm)}(r)$
in the perturbed theory with $\lam=\pm |\lam|$, respectively,
should be related by
analytic continuation $r^y \to -r^y$, up to the above non-analytic term.
This condition, implying the relation
\eqn\anpm{ a_n^{(+)}~=~(-1)^n a_n^{(-)} }
between the corresponding CPT coefficients, already
imposes strong constraints on scaling functions as
conjectured within the TBA approach.
The ultimate tests are made by comparing the explicit values of the
(regularized) CPT coefficients, eq.~(3.6), with the ones obtained from
the TBA results. Unfortunately, it is technically very difficult in general to
carry out the CPT calculations using (3.6) for more than the first one
or two leading coefficients, due to the lack of convenient representations
for the critical correlators involved;\foot{
See, however, the generically more powerful
method of calculating the $a_n$ (numerically)
based on {\it Hamiltonian} CPT, which is
discussed in \rourv. \nl} within the TBA approach, on the
other hand, until now it has been possible to obtain the expansion
coefficients only numerically (to very high accuracy, for the leading
ones).
Though
limited, these computations proved sufficient to provide
highly nontrivial    consistency checks on the
CPT and TBA results,
 and as a byproduct accurate values for $\kappa$
of \kappadef ~were obtained (Table~1 summarizes the
results \rRSOStba\rmlesstba\rourv ~for the theories relevant to this paper;
interestingly, the choice of mass scales in $A_m^{(\pm)}$ that we implicitly
made by taking the nonzero $\nu_a(\tht)$ in the TBA systems to be exactly
$\cosh\tht$ and ${1\over 2}e^{\pm\tht}$, respectively, leads to the same
$\kappa_m$~\rmlesstba\rourv\
in the two theories
within the numerical
accuracy).

\setbox\strutbox=\hbox{\vrule height12pt depth5pt width0pt}
\def\tablerule{\noalign{\hrule}}
\def\strut{\relax\ifmmode\copy\strutbox\else\unhcopy\strutbox\fi}

\vskip 6mm

\centerline{\vbox{\halign{&#\vrule&\strut$~#~$&
       #\vrule&\strut$~#~$&
       #\vrule\cr\tablerule
& ~~~m~~~ && \kappa_m               & \cr \tablerule\tablerule
& 3 && \hfill {1\over 2\pi}\hfill      & \cr \tablerule
& 4 && ~~~0.148695516112(3)~~~\hfill  & \cr \tablerule
& 5 && ~~~0.130234474(2)\hfill     & \cr \tablerule
& 6 && ~~~0.11334655(2)\hfill      & \cr \tablerule
& 7 && ~~~0.099267(2)\hfill        & \cr \tablerule}} }
\noindent {\bf Table~1:} $\kappa$, eq.~\kappadef,~in the first few models
$A_m^{(\pm)}$. The number in parenthesis is the estimated error in the last
 digit given.
(See (A.9) for a conjectured exact expression for $\kappa_5$.)
\vskip 6mm

The analysis of the IR limit $r\to\infty$ showed~\rmlesstba\rourv\ that
$e(A_m^{(+)},(p,p)|\infty) = d_{p,p}^{{\scsc (m-1)}}-{c_{m-1}\over 12}$
(with $p=1$ for all $m\ge 4$, as well as $p=2,{m\over 2}$ for even $m\ge 4$),
exhibiting the flows of the fields $\phi_{p,p}$ in question in $A_m$
to the fields with the same Kac labels in $A_{m-1}$. These flows of fields are
expected from the LG and perturbative RG analyses of
$A_m^{(+)}$.
More generally, these methods predict~\rZamflow\rLudCar\ for any $m\ge 4$
\eqn\Aphiflow{ \eqalign{
   \phi_{p,p} &~\to~ \phi_{p,p} ~~~~~~~~~~{\rm for} ~~p=1,\ldots,m-2\cr
   \phi_{m-1,m-1}  
                                       &~\to~ \phi_{2,1}     \cr
   \phi_{p+1,p}   &~\to~ \phi_{p+2,p+1} ~~~~{\rm for} ~~p=1,\ldots,m-4~,\cr} }
and in addition the perturbing field $\phi_{1,3}$ is expected~\rZamflow\ to
flow to the irrelevant field $\phi_{3,1}$
($T\bar{T}$ in the $m=4$ case~\rKMS) in $A_{m-1}$. The latter observation
lets one hope that   it is possible to
describe the large $r$ behaviour of $A_m^{(+)}$ using IR-CPT,
{\it i.e.}~perturbation theory around the IR CFT $A_{m-1}$
with the {\it irrelevant} perturbation $\phi_{3,1}$
(presumably together with an infinite series of more irrelevant fields).
 The large $r$ expansions of scaling functions
obtained using such IR-CPT are
expected to be only asymptotic, in contrast to
the UV-CPT expansions around $r=0$,
which most probably have  a nonzero radius of convergence~\rZamtba\rouriii.
Nevertheless, successful comparison~\rmlesstba\rourv\ of the leading terms in
these asymptotic expansions with the large $r$ behaviour of the conjectured
exact results from the TBA approach provides further evidence to the
correctness of the latter.

In~\rourv ~we noticed (but at that time did not understand) the following
interesting fact concerning the TBA systems
obtained from those of the ground state energy $e(A_m^{(\pm)},(1,1)|r)$ (see
above) with $m\ge 5$ {\it odd} by just changing the type to
{}~$t=(-1,\ldots,-1)$. Denote the corresponding solutions (temporarily) by
$e^{(\pm)}(m|r)$. Analytic calculation of the UV limit gives
{}~$e^{(\pm)}(m|0)=d_{{m+1\over 2},{m+1\over 2}}^{(m)}-{c_m\over 12}$.
This suggests to identify $e^{(\pm)}(m|r)$ as
$e(A_m^{(\pm)},({m+1\over 2},{m+1\over 2})|r)$.
However, for the `$+$'-case this identification turns
out to contradict the LG prediction
of \Aphiflow, since
\eqn\IRlim{ e^{(+)}(m|\infty)~=~d_{{\sc {m-1\over 2},{m-1\over 2}}}^{(m-1)}
           -{c_{m-1}\over 12}~. }
Furthermore, numerically solving the equations for
$e^{(\pm)}(m|r)$ for $m=5,7$ we determined the small $r^{y_m}$ expansion
coefficients $a_n^{(\pm)}(m)$ (see eq.~\eofrexp ) which we present in Table~2.
The relation \anpm ~predicted by CPT for
$a_n(A_m^{(\pm)},({m+1\over 2},{m+1\over 2}))$ is {\it not} satisfied,
though the equality ~$a_1^{(+)}(m)=a_1^{(-)}(m)$ ~(no minus sign!) is notable.
We also noticed in \rourv ~that in absolute value these equal coefficients
agree with the CPT prediction (eq.~(3.6), where the value of $\kappa_m$
(see Table~1) that must be used for this comparison
was  obtained from the analysis of the ground state scaling function.

\vskip 9mm
\setbox\strutbox=\hbox{\vrule height15pt depth5pt width0pt}
\centerline{\vbox{\halign{&#\vrule&\strut$~#~$&
               #\vrule&\strut$~#~$&
               #\vrule&\strut$~#~$&
               #\vrule&\strut$~#~$&
               #\vrule&\strut$~#~$&
               #\vrule\cr\tablerule
& ~n~ && ~a_n^{(+)}(5)~ && ~a_n^{(-)}(5)~
&& ~~a_n^{(+)}(7)~ && ~~a_n^{(-)}(7)~& \cr \tablerule\tablerule
& 1 && ~-0.056643444(3)\hfill && ~-0.056643449(4)\hfill && ~-0.042641(1)\hfill
                                 && ~-0.042640(1)\hfill & \cr \tablerule
& 2 && ~~~0.0484328(4)\hfill && ~-0.0012546(5)\hfill && ~~~0.03918(1)\hfill
                                 && ~-0.01743(2)\hfill  & \cr \tablerule
& 3 && ~-0.020718(4)\hfill  && ~~~0.01173(1)\hfill && ~-0.0366(2)\hfill
                                 && ~~~0.0100(5)\hfill   & \cr \tablerule
& 4 && ~~~0.00140(4)\hfill   && ~~~0.0006(1)\hfill && ~~~0.017(2)\hfill
                                 && ~~~0.008(2)\hfill    & \cr \tablerule}} }
\vskip 1mm
\noindent{\bf Table~2:}
The first few coefficients in the (regular part of the) small
$r$ expansion of the
functions $e^{(\pm)}(m|r)$, $m=5,7$, conjectured to describe scaling
functions of the excitations specified in eq.~(2.9).

\vskip 7mm

We here propose an ``explanation'' for the above observations.
We conjecture that for $m$ {\it odd} the correct identification of the above
functions $e^{(\pm)}(m|r)$ obtained in the TBA approach (namely the solutions
of \eofr --\epseq ~with $N=m-2$, $t=(-1,\ldots,-1)$,
$K(\tht)=I^{(m-2)}/\cosh\tht$, and
{}~$\nu(\tht)=(\cosh\tht,0,\ldots,0)$~ for $e^{(-)}(m|r)$,
{}~$\nu(\tht)=(\shalf e^\tht,0,\ldots,0,\shalf e^{-\tht})$~ for $e^{(+)}(m|r)$)
correspond to the following scaling functions:
\eqn\conj{ \eqalign{
     e^{(-)}(m|r) &~=~ e(A_m^{(-)},({\textstyle {m+1\over 2}},
                                      {\textstyle {m+1\over 2}})|r) \cr
     e^{(+)}(m|r) &~=~ e(D_m^{(+)},({\textstyle {m+1\over 2},
                                          {m+1\over 2}})^- |r)~.\cr} }
[The superscript `$-$' on the Kac label $({m+1\over 2},{m+1\over 2})$
on the second line indicates that the corresponding UV conformal state
is created by the $\ZZ_2$-{\it odd} spinless primary field
$\phi_{{m+1\over 2},{m+1\over 2}}$ in $D_m$.
This is crucial when $m\equiv 1~({\rm mod}~4)$, in which case $D_m$ contains
two copies of $\phi_{{m+1\over 2},{m+1\over 2}}$.]

In the following we will present evidence for the conjecture \conj.
If true, one concludes from \IRlim\  and the fact that the perturbing
field does not break the $\ZZ_2$ symmetry, that
\eqn\Dphiflow{ \eqalign{
    \phi_{3,3}^- ~\to~ \phi_{2,2} ~~~~~~~~~~~~~~& {\rm in}~~~
                    D_5^{(+)}: D_5 \to A_4 \cr
    \phi_{{m+1\over 2},{m+1\over 2}} ~\to~ \phi_{{m-1\over 2},{m-1\over 2}}^-
  ~~~~~~& {\rm in}~~~D_m^{(+)}: D_m \to D_{m-1}~~~~~(m=7,11,15,\ldots)\cr
    \phi_{{m+1\over 2},{m+1\over 2}}^- ~\to~ \phi_{{m-1\over 2},{m-1\over 2}}
 ~~~~~~& {\rm in}~~~ D_m^{(+)}: D_m \to
                                        X_{m-1} ~~~~~(m=9,13,17,\ldots)~,
     \cr} }
which is our concrete evidence for \Dflow ~when $m$ is odd.
(Using IR-CPT we will argue in sect.~3.4 that $X_{m-1}$ in the last
line of \Dphiflow\ is in fact $D_{m-1}$. See also sect.~3.3.)
Based on CPT arguments, we also claim that for the $\ZZ_2$-{\it even} spinless
primary fields $\phi_{p,q}$ in the model $D_m$, one has
\eqn\Dphieven{ e(D_m^{(\pm)},(p,q)|r) ~=~ e(A_m^{(\pm)},(p,q)|r)~
      ~~~~~~~~(\phi_{p,q}~ {\rm even})~.}
Similarly, for the $\ZZ_2$-odd fields $\phi_{{m\over 2},q}^-$
{}~($q=1,\ldots,{m\over 2}$) in $D_m$ with $m$ even
\eqn\Dphiodd{
   e(D_m^{(\pm)},({\textstyle {m\over 2}},q)^-|r) ~=~
       e(A_m^{(\pm)},({\textstyle {m\over 2}},q)|r)
        ~~~~~~~~~~(m=6,8,10,\ldots)~.}
Assuming the LG/perturbative-RG predictions \Aphiflow\
(we do not have any TBA results for most of the scaling functions involved)
we conclude that the fields in $D_m$ corresponding to the
scaling functions in \Dphieven --\Dphiodd ~flow under the
$\phi_{1,3}$-perturbation to the fields
(possibly non-primary) with the same Kac labels as in the $A$-flows,
with the $\ZZ_2$ symmetry resolving possible ambiguities due to doubling
of fields in the $D$-models.
To complete the description of flows of relevant fields in $D_m$ that
remain relevant in the IR CFT $D_{m-1}$, it is natural to guess that
$\phi_{{m+3\over 2},{m+1\over 2}}^- \to \phi_{{m-1\over 2},{m-3\over 2}}$
for $m=5,9,13,\ldots$, with the `$-$' superscript moved over to the IR field
when $m=7,11,15,\ldots$.
These observations, and the fact that they provide support to our main
claim \Dflow, will become clearer once we discuss the $D$-models
and their perturbations in more detail.

\newsec{CPT analysis of the theories $D_m^{(\pm)}$}

\subsec{The CFTs $D_m$}

The field content of the models $D_m$  is encoded in their MIPFs, written
in terms of Virasoro characters~\rCarMI\rGep\rCIZ.
We have to distinguish between four
cases:\nl
\noindent
(i) $m=4\rho+1$ ~($\rho\ge 1$):
\eqn\ZDi{ \eqalign{
  Z~=~ &\sum_{p=1}^{4\rho} ~\sum_{q~{\rm odd}=1}^{2\rho-1} |\chi_{p,q}|^2
      + \sum_{p=1}^{2\rho} |\chi_{p,2\rho+1}|^2   \cr
     &+ \sum_{p=1}^{2\rho} |\chi_{p,2\rho+1}|^2
      + \sum_{p=1}^{4\rho} ~\sum_{q~{\rm odd}=1}^{2\rho-1}
                            \chi_{p,q} ~\chi_{4\rho+1-p,q}^* ~. \cr} }
\noindent
(ii) $m=4\rho+2$ ~($\rho\ge 1$):
\eqn\ZDii{ \eqalign{
  Z~=~
      &\sum_{p~{\rm odd}=1}^{2\rho-1} ~\sum_{q=1}^{4\rho+2} |\chi_{p,q}|^2
            + \sum_{q=1}^{2\rho+1} |\chi_{2\rho+1,q}|^2   \cr
           &+ \sum_{q=1}^{2\rho+1} |\chi_{2\rho+1,q}|^2
    + \sum_{p~{\rm odd}=1}^{2\rho-1} ~\sum_{q=1}^{4\rho+2}
                    \chi_{p,q} ~\chi_{p,4\rho+3-q}^* ~. \cr} }
\noindent
(iii) $m=4\rho-1$ ~($\rho\ge 2$):
\eqn\ZDiii{ \eqalign{
  Z~=~ &\sum_{p=1}^{4\rho-2} ~\sum_{q~{\rm odd}=1}^{2\rho-1} |\chi_{p,q}|^2 \cr
     & +\sum_{p=1}^{2\rho-1} |\chi_{p,2\rho}|^2
       + \sum_{p=1}^{4\rho-2} ~\sum_{q~{\rm even}=2}^{2\rho-2}
                    \chi_{p,q}~ \chi_{p,4\rho-q}^* ~. \cr} }
\noindent
(iv) $m=4\rho$ ~($\rho\ge 2$):
\eqn\ZDiv{ \eqalign{
  Z~=~
  &\sum_{p~{\rm odd}=1}^{2\rho-1} ~\sum_{q=1}^{4\rho} |\chi_{p,q}|^2 \cr
     &+ \sum_{q=1}^{2\rho} |\chi_{2\rho,q}|^2
   + \sum_{p~{\rm even}=2}^{2\rho-2} ~\sum_{q=1}^{4\rho}
                    \chi_{p,q}~ \chi_{4\rho-p,q}^* ~. \cr} }
This (not always economic) way of writing the MIPFs makes clear the following
facts. Each term $|\chi_{p,q}|^2$ in $Z$ corresponds to
(the conformal family of) a spinless
primary field $\phi_{p,q}$ in the model, whereas a term
$\chi_{p,q}\chi_{\bar{p},\bar{q}}^*$, always with $(\bar{p},\bar{q})$
different from $(p,q)\equiv (m-p,m+1-q)$, corresponds to a primary field
 $\phi_{p,q;\bar{p},\bar{q}}$ with nonzero spin.
In addition, the fields corresponding to
the terms on the first (second) line in each case are $\ZZ_2$-even(odd) \rGep.
Note that the models $A_m$ and $D_m$ ($m\ge 5$) have the same
$\ZZ_2$-even primary fields. On the other hand, the $\ZZ_2$-odd (spinless)
primary fields in $A_m$ are ``replaced'' by primary fields with nonzero spin
in $D_m$, which also contains $[{m\over 2}]$ ``extra''
odd primary fields which are spinless
(and lead to a doubling of fields iff $m\equiv 1~{\rm or}~2~({\rm mod}~4)$).
 In particular,
all the relevant primary fields in $D_m$ are spinless,\foot{
A notable exception is the 3-state Potts model $D_5$ that contains
two relevant ($d={9\over 5}$) primary fields with nonzero spin:
$\phi_{2,1;3,1}$ and $\phi_{3,1;2,1}$. \nl}
and there are $m-2$ such even fields (excluding the
trivial identity field $\phi_{1,1}$) and two odd ones;
in $A_m$, on the other hand, there are $2(m-2)$ nontrivial
relevant (spinless, primary) fields, half of them even and half odd.

An important ingredient in the CPT analysis below is the structure of the
operator algebra of the models $D_m$. The fusion rules for these
models were given explicitly in \rGep. They are essentially the
$A$-series fusion rules (applied separately to the left and right Kac label
pairs $(p,q)$ and $(\bar{p},\bar{q})$ appearing in
 \ZDi--\ZDiv) intersected with the $\ZZ_2$-symmetry selection rules.
On the other hand,
the problem of finding the (nonvanishing) operator product expansion
coefficients (OPECs) is much  more involved~\rPetk\ than in the
corresponding $A$-models.
This is the case for CFTs with non-diagonal MIPFs in general~\rFuKl.
The main subtlety lies in the signs of certain OPECs in the non-diagonal
models. What is clear~\rPetk\rFuKl, however, is that if the fusion
of two spinless primary fields in the non-diagonal
model gives rise to primary fields that are all spinless, then the OPECs
involved are identical to those in the diagonal model.\foot{More generally,
if the correlator of some fields in conformal families whose ancestors are
all spinless primary fields gets contributions from intermediate fields
(through ``factorization'') all of whose ancestors are again spinless,
then this correlator is the same as in the corresponding CFT with diagonal
MIPF. This follows from the uniqueness of the solution to the
monodromy problem~\rFuKl. \nl}
For $D_m$, it is known~\rPetk\
that the nonvanishing OPECs (of primary fields of definite $\ZZ_2$ parity)
satisfy
\eqn\ADOPEC{ \left(\IC_{(p_1,q_1;\bar{p}_1,\bar{q}_1)
                        (p_2,q_2;\bar{p}_2,\bar{q}_2)
                        (p_3,q_3;\bar{p}_3,\bar{q}_3)}^{(D_m)} \right)^2 =
                    ~\IC_{(p_1,q_1)(p_2,q_2)(p_3,q_3)}^{(A_m)}
  ~ \IC_{(\bar{p}_1,\bar{q}_1)(\bar{p}_2,\bar{q}_2)(\bar{p}_3,
     \bar{q}_3)}^{(A_m)}~,}
where the OPECs on the right-hand side are those of the  primary fields
in $A_m$~\rDotFatiii.
Now recall that the primary fields in $A_m$ are normalized, up to signs,
by requiring
that $\langle \phi_{p_1,q_1}(\infty,\infty)~\phi_{p_2,q_2}(0,0)\rangle
= \delta_{p_1 p_2} \delta_{q_1 q_2}$ (here $1\le q_i \le p_i \le m-1$).
The signs can be further fixed so that all the OPECs in $A_m$ are
non-negative~\rDotFatiii. Hence eq.~\ADOPEC\
gives us the OPECs in $D_m$ up to signs, which turn out to be crucial for
the discussion below.

\subsec{UV-CPT for $D_m^{(\pm)}$}

We are now ready to use CPT to study the theories
$D_m^{(\pm)}$, trying to support
the validity of eqs.~\conj--\Dphiodd\ that are our evidence for the main
claim \Dflow. For the scaling functions $e(X_m^{(\pm)},(p,q)|r)$, CPT
based on \pCFT~gives the small $r$ expansion \eofrexp~with the CPT coefficients
\eqn\anCPT{ \eqalign{  a_n(& X_m^{(\pm)},(p,q)) ~=~
   -(2\pi)^{1-y_m}~(\mp \kappa_m)^n  \cr
   &\times \int \prod_{j=1}^{n-1} {d^2z_j \over (2\pi|z_j|)^{y_m}}~
     \langle \phi_{p,q}(\infty,\infty) ~\phi_{1,3}(1,1) \prod_{j=1}^{n-1}
    \phi_{1,3}(z_j,\bar{z}_j) ~\phi_{p,q}(0,0)\rangle_{{\rm conn}} ~.\cr} }
The correlators here are connected
(with respect to the ``in- and out-states'' created by $\phi_{p,q}$)
$(n+2)$-point functions in the CFT $X_m$ on the plane.
[On the r.h.s.~we suppressed the factor $\langle \phi_{p,q}(\infty,\infty)
{}~\phi_{p,q}(0,0)\rangle^{-1}$, which is set to 1. In particular, we will
assume this standard CFT normalization for the combinations $\phi_{p,q}^\pm$
of definite $\ZZ_2$-parity in the  relevant
$D$-models. Since in the perturbations of these latter models
(like in all the other $A^{(\pm)}$ and $D^{(\pm)}$ theories)
the $\ZZ_2$ charge is a good quantum number also away from criticality,
it makes sense to consider states
in the perturbed theory whose UV limits correspond to $\phi_{p,q}^+$ and
$\phi_{p,q}^-$.]
The problem of (UV and IR) regularization of such integrated correlators
was discussed at length in~\rourv\
(see also the appendix).

Of particular importance to us will be the first CPT coefficient, which is
never divergent, given simply in terms of conformal OPECs by
\eqn\aoneCPT{ a_1(X_m^{(\pm)},(p,q)) ~=~ \pm (2\pi)^{1-y_m}~\kappa_m
    ~~\IC_{(p,q)(1,3)(p,q)}^{(X_m)}~.}

Note that it always vanishes for the ground state
(in the unitary theories under consideration)  corresponding
to $(p,q)=(1,1)$.

\vskip 3mm

Let us now prove eq.~\Dphieven. This equation is a consequence of the fact that
all spinfull primary
fields in the $D$-models are $\ZZ_2$-odd and therefore  cannot
appear in the fusion of two $\ZZ_2$-even fields.
It is therefore clear (cf.~footnote~7)
that the CPT coefficients involved are identical, including signs, in
$D_m^{(\pm)}$ and $A_m^{(\pm)}$. Analytic continuation of the small
$r$ CPT expansion to all $r$ proves \Dphieven.
Actually
we see from the above argument, which did not use the fact that the
$\ZZ_2$-even fields $\phi_{p,q}$ involved are primary, that eq.~\Dphieven
{}~generalizes to the statement that the whole $\ZZ_2$-even sectors of
the finite-volume spectra of $D_m^{(\pm)}$ and $A_m^{(\pm)}$ are identical.

Similarly, \Dphiodd ~follows from the fact that in the fusion
(in $D_m$, $m$ even)
\eqn\DevenFusion{[\phi_{1,3}] ~\times~ [\phi^-_{{m\over 2}, q}] ~=~
        [\phi^-_{{m\over 2}, q-2}] ~+~ [\phi^-_{{m\over 2}, q}] ~+~
            [\phi^-_{{m\over 2}, q+2}]  }
no primary fields of nonzero spin  can appear on the r.h.s.~(there
are no  such fields for which the first entries of the
left and right Kac label pairs are both
$\sfrac{m}{2}$). Again, this implies that the relevant CPT coefficients in
$A^{(\pm)}_m$ and $D^{(\pm)}_m$  are identical
(this is true also for all the scaling functions corresponding to
descendants of the primary fields $\phi_{{m\over 2},q}^-$);
the fact that
$\phi_{{m\over 2},q}$ is $\ZZ_2$-even in $A_m$ if $m\equiv 2~({\rm mod}~4)$
does not matter. Note that in this latter case \Dphieven --\Dphiodd ~imply,
in particular,
that the $\ZZ_2$-doublet $\phi^{\pm}_{{m\over 2},{m\over 2}}$ in $D_m$
flows to the doublet with the same Kac label in $D_{m-1}$.

\vskip 2mm

Now to \conj--\Dphiflow, our most interesting claim.
Note first of all that in
the conjectured flow $\phi_{{m+1\over 2},{m+1\over 2}}
\to \phi_{{m-1\over 2},{m-1\over 2}}$
 (with appropriate `$-$' superscripts as in \Dphiflow)
in $D_m \to D_{m-1}$, $m$ odd,
the field becomes more relevant during the flow, {\it i.e.}~its scaling
dimension decreases:
\eqn\dflow{d^{{\sc (m-1)}}_{{m-1\over 2},{m-1\over 2}} ~-~ d^{{\sc (m)}}_
   {{m+1\over 2},{m+1\over 2}} ~=~
  -{{1}\over{4m}}~ {{m^2+3}\over{m^2-1}} ~<~ 0 ~.}
This is in contrast to all the known flows of fields in the $A$-series, where
fields become less relevant (the fact that this feature of the $A$-series
flows, reminiscent of the decrease of the central charge along unitary
RG trajectories~\rZamflow, is not universal, was already noted in \rKMS ).
The qualitative difference can be seen already in the first order of the
small $r$ expansion of the relevant scaling function:
Remember our empirical observation concerning the first expansion
coefficients of the ``scaling functions'' $e^{(\pm)}(m|r)$
of the type $t=(-1,\ldots,-1)$ TBA systems with $m$ odd.
Assuming \conj\ we can now state it equivalently as
\eqn\aAD{ \eqalign{
 a_1(D^{(+)}_m, ({\textstyle {m+1\over 2},{m+1\over 2}})^-)
      & ~=~ +~a_1(A^{(-)}_m, ({\textstyle {m+1\over 2},{m+1\over 2}}))~
         ~~~~~~~(m~{\rm odd})
                                                                         \cr
      & ~=~ -~a_1(A^{(+)}_m, ({\textstyle {m+1\over 2},{m+1\over 2}}))~,\cr}}
where the `$-$' superscript on the l.h.s.~can be ignored in the case
$m\equiv 3~({\rm mod}~4)$.
Actually we found~\rourv ~the first equality (the second
is just a particular case of \anpm )
 only for $m=5,7$, but in the meantime also have checked it
for $m=9, 11$, and believe that it holds for all odd $m$.

\vskip 3mm

What is the origin of this difference between the flows in the $D$ and the $A$
series from the point of view of CPT? We claim that for
$m\equiv 3~({\rm mod}~4)$ the crucial fact, proving \aAD, is
\eqn\ADiii{ \IC^{(D_m)}_{({m+1\over 2},{m+1\over 2})(1,3)
              ({m+1\over 2},{m+1\over 2})} ~=~
           -~\IC^{(A_m)}_{({m+1\over 2},{m+1\over 2})(1,3)
              ({m+1\over 2},{m+1\over 2})} ~,~~~~m\equiv 3~({\rm mod}~4)~.}
The crucial sign difference is consistent with the fact that
now spinfull fields appear in the fusion
{}~$[\phi_{1,3}] \times [\phi_{{m+1\over 2},{m+1\over 2}}]$.\foot{Unless
we misunderstand the results of~\rPetk, our \ADiii\ disagrees with them
when $m\equiv 3~({\rm mod}~8)$. The same remark applies to our
(3.12), now with $m\equiv 1~({\rm mod}~8)$. \nl}
Note that sign differences between OPECs like the one in~\ADiii\
have a less trivial
effect on higher CPT coefficients, so that for $n>1$ we
do not expect any simple relations like~\aAD.
This is consistent with our numerical results --- cf.~Table~2.

\vskip 3mm

For $m\equiv 1~({\rm mod}~4)$, on the other hand, the crucial fact is,
we claim,
\eqn\ADi{\eqalign{
  \IC^{(D_m)}_{({m+1\over 2},{m+1\over 2})^-(1,3)
                               ({m+1\over 2},{m+1\over 2})^-}
 &= -~\IC^{(D_m)}_{({m+1\over 2},{m+1\over 2})^+ (1,3)
                               ({m+1\over 2},{m+1\over 2})^+}~,
       ~m\equiv 1~({\rm mod}~4)
             \cr
 &= -~\IC^{(A_m)}_{({m+1\over 2},{m+1\over 2}) (1,3)
                               ({m+1\over 2},{m+1\over 2})}~ ~.\cr}}
This allows the scaling dimension to {\it decrease} in the flow
$\phi^-_{{m+1\over 2},{m+1\over 2}} \to \phi_{{m-1\over 2},{m-1\over 2}}$
in $D_m^{(+)}$, whereas in the
flow $\phi^+_{{m+1\over 2},{m+1\over 2}} \to \phi_{{m+1\over 2},{m+1\over 2}}$
the scaling dimension increases, as in the corresponding $A$-series flow.

For $m=5$ we can use the $\ZZ_3$ symmetry of the 3-state Potts model $D_5$
to easily prove the first equality in~\ADi\
(note that the second equality holds in general, according to our remarks
in sect.~3.1).  Under this ${\bf Z}_3$-symmetry
$\phi_{3,3} \equiv \sfrac{1}{\sqrt{2}}(\phi^+_{3,3} + i \phi^-_{3,3})$ and
$\phi^{\ast}_{3,3} \equiv \sfrac{1}{\sqrt{2}} (\phi^+_{3,3} - i \phi^-_{3,3})$
form a doublet of oppositely
charged fields.
Recall that the perturbation is by
$\phi^+_{1,3} = \sfrac{1}{\sqrt{2}}(\phi_{1,3} + \phi^{\ast}_{1,3})$,
which implies
\eqn\CDv{\IC^{(D_5)}_{(3,3)^\pm (1,3)^+ (3,3)^\pm}  =~
   \pm{\textstyle {1\over 2\sqrt2}}~\IC^{(D_5)}_{\phi_{3,3}\pm\phi_{3,3}^\ast,
                    ~\phi_{1,3}+\phi_{1,3}^\ast,~\phi_{3,3}\pm\phi_{3,3}^\ast}
 =~ \pm {1\over \sqrt{2}}~\IC^{(D_5)}_{\phi_{3,3},~\phi_{1,3},~\phi_{3,3}} ~,}
where we used the linearity of the OPECs
and in the last step also the conservation of the $\ZZ_3$ charge
and the ${\bf Z}_2$ symmetry
(we choose to assign the same $\ZZ_3$ charge to $\phi_{3,3}$ and $\phi_{1,3}$).
This proves \ADi ~for $m=5$.

[It is interesting to explicitly see how different the consequences of the
$\ZZ_3$ symmetry are in the theory $D_6^{(\pm)}$.
($D_6$, the tricritical 3-state Potts CFT~\rCIZ, and $D_5$ are the only
$\ZZ_3$-symmetric models among the $X_m$.)
  First, note that there is a single $\ZZ_3$- (and $\ZZ_2$-)neutral
field $\phi_{1,3}$ in $D_6$,
in contrast to $D_5$, so that
both the $\ZZ_2$ and the $\ZZ_3$ symmetries are
preserved in the perturbed theories $D_6^{(\pm)}$.
(This fact by itself shows that $D_6^{(+)}$ cannot possibly flow
to $A_5$, as $A_5$ is not $\ZZ_3$-symmetric.)
Now consider the first CPT coefficients relevant for the
flow of the doublet $\phi_{3,3}^\pm \to \phi_{3,3}^\pm$ in $D_6^{(+)}$.
Here, in contradistinction to \CDv, conservation of the $\ZZ_3$ charge gives
\eqn\CDvi{\IC^{(D_6)}_{(3,3)^\pm (1,3) (3,3)^\pm} ~=~
       \pm{\textstyle {1\over 2}}~\IC^{(D_6)}_{\phi_{3,3}\pm\phi_{3,3}^\ast,
                          ~\phi_{1,3},~\phi_{3,3}\pm\phi_{3,3}^\ast} ~=~
   ~\IC^{(D_6)}_{\phi_{3,3},~\phi_{1,3},~\phi_{3,3}^\ast} ~, }
implying that $a_1(D_6^{(+)},(3,3)^+)=a_1(D_6^{(+)},(3,3)^-)$ as
expected from  other considerations discussed earlier.]

\subsec{RG-improved CPT}

To gain {\it direct} information about the IR fixed point from standard UV-CPT
on the cylinder, as discussed in the last subsection, would require analytic
continuation of the small $r$ expansion to large $r$. This is generically not
possible in practice.
However, one can sometimes
directly ``see'' the IR
fixed point by using the RG to (partially) sum up perturbation theory. This
assumes, of course, that one knows the $\beta$-function quite accurately up
to its IR zero. In the present context this restricts the applicability
of this approach to large $m$, where the IR fixed point is close to the UV one
(as measured by the {\it renormalized} coupling,  not the bare coupling
$\lambda$ of sect.~3.2
in terms of which the IR fixed point is at $\lambda=\infty$).
This method has been discussed in detail in the
literature~\rZamflow\rLudCar\rCapLat, so there is no need to review it.
We just summarize the results relevant for us.

Consider the RG flow induced by perturbing a CFT of central charge $c$ by a
slightly relevant ($y \ll 1$)  operator $\phi$,
which up to irrelevant operators
closes on itself and the identity operator under repeated fusions.
The central charge $c'$ at the new IR fixed point (attained by choosing the
appropriate sign for the perturbing term) then satisfies
\eqn\cflow{ c' - c ~=~ -{y^3 \over (\IC_{\phi\phi\phi})^2} ~+~ {\cal O}(y^4)~.}
 For a field $\phi_{\alpha}$ which to leading order does not mix with
any other field (like the fields $\phi_{p,p}$ on the ``diagonal''
of the Kac table of a minimal model) the change in scaling dimension is
\eqn\dflowii{d'_\alpha - d_\alpha ~=~
 {2~\IC_{\phi_{\alpha}\phi\phi_{\alpha}}
    \over \IC_{\phi\phi\phi} }~y  ~+~ {\cal O}(y^2)~.}

In the case we are interested in $\phi=\phi_{1,3}$, ~$y=\sfrac{4}{m+1}$,
{}~$\IC_{\phi\phi\phi}=\sfrac{4}{\sqrt{3}} + {\cal O}(\sfrac{1}{m})$
{}~\rZamflow\rLudCar,
  so that $c' -c =-\sfrac{12}{m^3} +
{\cal O}(\sfrac{1}{m^4})$, consistent with a flow $X_m \to X_{m-1}$.
 For $\phi_\alpha = \phi_{p,p}$ with $p \leq [\sfrac{m}{2}]$ odd the OPECs
{}~$\IC_{\phi_{\alpha}\phi\phi_{\alpha}} = \sfrac{p^2 -1}{2 \sqrt{3} m^2} +
{\cal O}(\sfrac{p^2}{m^3})$ ~\rZamflow\
are identical in $A_m$ and
$D_m$, showing the existence of flows $\phi_{p,p} \to \phi_{p,p}$ in
$X_m \to X_{m-1}$. These flows do not allow us to directly distinguish
between, say, $D_m \to D_{m-1}$ and $D_m \to A_{m-1}$ (although it is
clearly hard to see where the fields $\phi_{p,p}$, $p$ even, in $A_{m-1}$
should come from!).

 For $m\equiv 3~({\rm mod}~4)$, however, the flow of the $\ZZ_2$-odd field
$\phi_\alpha = \phi_{{m+1\over 2},{m+1\over 2}}$ does enable us to directly
distinguish the two possibilities for flows starting from $D_m$.
Depending on the
sign of ~$\IC_{\phi_{\alpha}\phi\phi_{\alpha}} = \pm (\sfrac{1}{8\sqrt{3} }
 +{\cal O}(\sfrac{1}{m}))$    we now have
\eqn\dflowiii{d'_\alpha - d_\alpha = \pm {1\over 4 m} ~+~
                                      {\cal O}({\textstyle {1 \over m^2}}) ~,}
indicating the flow of scaling dimensions
$d^{{\scsc (m)}}_{{m+1\over 2},{m+1\over 2}} \to
  d^{{\scsc (m-1)}}_{{m+1\over 2},{m+1\over 2}}$    ~or~
$d^{{\scsc (m)}}_{{m+1\over 2},{m+1\over 2}} \break \to
  d^{{\scsc (m-1)}}_{{m-1\over 2},{m-1\over 2}}$,
respectively.
According to our claim \ADiii\  the second possibility holds for
$D_m^{(+)}$, which
is only possible if $D_m \to D_{m-1}$ (because of the $\ZZ_2$ symmetry, as
there is a single field $\phi_{{m-1\over 2},{m-1\over 2}}$ --- which is
$\ZZ_2$-even --- in $A_{m-1}$).

For $m\equiv 1~({\rm mod}~4)$ the analogous sign difference \ADi\ again
indicates the flow
$\phi^-_{{m+1\over 2},{m+1\over 2}} \to \phi_{{m-1\over 2},{m-1\over 2}}$
in $D_m^{(+)}$. But now, unfortunately, this flow alone does not
allow us to decisively conclude that the IR CFT is $D_{m-1}$. In fact, for
$m=5$ we know this cannot be the case.
To show that $D_m \to A_{m-1}$ is not possible
for $m=9,13,17,\ldots$ requires the study of flows of other fields,
not on the ``diagonal'' of the Kac table, but we will not pursue such
studies here. Intuitively, the fact that there are about half as many
relevant fields in $D_m$ than in $A_{m-1}$, for $m\gg 1$, makes the flow
between these two theories highly implausible. A more explicit version of this
argument will be offered in sect.~3.4.

 For $m$ even the OPEC ~$\IC_{\phi_{\alpha}\phi\phi_{\alpha}}$,
$\phi_{\alpha} = \phi_{{m\over 2},{m\over 2}}$,
in $D_m$ (with
superscripts `$\pm$' for $m\equiv 2~({\rm mod}~4)$)
is identical to the corresponding one in $A_m$.
Therefore $\phi^{{\sc \pm}}_{{m\over 2},{m\over 2}}$ in $D_m$,
$m\equiv 2~({\rm mod}~4)$, must flow to fields
$\phi^{{\sc \pm}}_{{m\over 2},{m\over 2}}$
in $X_{m-1}$, which is only possible if $X_{m-1}=D_{m-1}$. For
$m\equiv 0~({\rm mod}~4)$ there is only one ($\ZZ_2$-odd)
field $\phi_{{m\over 2},{m\over 2}}$, and so
this argument cannot be applied.
As in the case of $m\equiv 1~({\rm mod}~4)$, it is necessary to study the flow
of other fields to conclude within the RG-improved CPT approach
that $D_m \to A_{m-1}$ is not possible.

\subsec{IR-CPT for $D_m^{(+)}$}

We finally discuss what can be learned about the  RG flows we are
considering from CPT around the IR fixed point. The basic idea
is that the action
\eqn\IRpCFT{  A_{A_{m-1}} + \int d^2 x ~[g~\phi_{3,1}(x) +~ \ldots~]~,
   ~~~~~~~~~m\ge 5 }
can be used to study perturbatively the IR asymptotics of the theory
$A_m^{(+)}$, in particular~\rmlesstba\rourv ~the large $r$ behaviour of the
energy scaling functions in this theory.
The leading nontrivial term in these scaling functions will generically be
proportional to $r^{-\sfrac{4}{m-1}}$, since
$d_{3,1}^{{\sc (m-1)}}=2+{4\over m-1}$.
The `$\ldots$' in the integrand refers
 to a presumably infinite series of fields
more irrelevant than $\phi_{3,1}$. (Note that the perturbation theory we are
considering here corresponds to a non-renormalizable interaction in standard
Lagrangian QFT.) Given the first (or perhaps more, see below) terms in
\IRpCFT, the infinite and finite parts of the counterterms written
as `$\ldots$' are determined, in principle,
by two conditons:
{}~a) That they make $E_0(R)$, say, finite order by order in perturbation
theory,
and ~b) that the presumably asymptotic large $r$ series is consistent with
the TBA and/or the (analytic continuation of the small $r$) UV-CPT series.
In~\rmlesstba\rourv ~it was shown that the first few terms (independent of
`$\ldots$') in the large $r$ expansion of  certain scaled energies for
$m=4,5,6,7$, calculated from \IRpCFT\ as in UV-CPT,
agree with those obtained in the TBA approach.

We will now examine what \IRpCFT ~and its generalization
to the $D$-flows can teach us. For $m\ge 6$ we should consider,
according to \Dflow\ and \Dphieven,
the action \IRpCFT ~with $A_{m-1}$ replaced by $D_{m-1}$
(and $\phi_{3,1}$ taken to be $\phi_{3,1}^+$ when $m=7$; note that in this
case the IR perturbation breaks the $\ZZ_3$ symmetry of $D_6$, which
explains the accidental --- from the point of view of UV-CPT --- appearance
of this symmetry at the IR limit of the theory $D_7^{(+)}$).\foot{
A similar situation occurs in the flows of $A_5$ and $D_5$ to $A_4$, the
(superconformal) tricritical Ising CFT. There the IR perturbation $\phi_{3,1}$,
the bottom rather than the top component of (the determinant of) the
super-stress-energy tensor, explicitly breaks the supersymmetry. \nl}
 The comparison between the ``backward flows''
induced in the $D$-series and those in the
$A$-series essentially parallels the previous discussion of UV-CPT.
Again, the conclusions are that flows of certain fields  are the same
in the two series (provided the IR perturbation is identical), whereas
for other flows ({\it e.g.}~\Dphiflow) the difference is consistent
with differences in the operator algebras of the models $D_{m-1}$ and
$A_{m-1}$.

The most interesting case is $m=5$, on which we now elaborate.
Here IR-CPT seems to present us with a little   puzzle:
If we are correct in our claim that $A_4$ is the common IR limit of both
$A_5^{(+)}$ and $D_5^{(+)}$, and moreover that the IR fixed point
is approached in both cases along the direction of $\phi_{3,1}$,
what is the difference in the IR perturbations of $A_4$ that leads to
the two different
``backward flows'' to $A_5$ and $D_5$?
{}~Recall that by UV-CPT we concluded that the $\ZZ_2$-even sector of the
spectrum of the two theories is identical, so that the IR limits
of UV fields in the conformal families of
$\phi_{p,1}$ ($p=1,2,3,4$) and $\phi_{p,3}^+$ ($p=1,3$) in $D_5$ and $A_5$
(the superscript `$+$' being redundant in the latter) is the same.
In particular, by \Aphiflow ~we have for the relevant $\ZZ_2$-even UV fields
$\phi_{1,1}\to \phi_{1,1}$,
$\phi_{3,3}^+\to \phi_{3,3}$,  $\phi_{2,1}\to \phi_{1,3}$, and
$\phi_{1,3}^+\to \phi_{3,1}$.
However,
the $\ZZ_2$-odd sector is necessarily different: Our main TBA
result \conj ~implies that $\phi_{3,3}^-\to \phi_{2,2}$ in $D_5^{(+)}$
whereas \Aphiflow ~predicts $\phi_{2,2}\to \phi_{2,2}$ in $A_5^{(+)}$;
\Aphiflow ~also predicts $\phi_{1,2}\to \phi_{2,1}$ in $A_5^{(+)}$,
and our guess for the
``corresponding'' flow in $D_5^{(+)}$ is $\phi_{1,3}^-\to \phi_{2,1}$.
We also believe that all the other $\ZZ_2$-odd fields in $D_5$, including
the primary spinfull ones, flow to descendants in the families $[\phi_{2,2}]$
and $[\phi_{2,1}]$ in $A_4$, the UV origin of these families in $A_5$
being $[\phi_{p,2}]$ ($p=1,2,3,4$).

Trying to understand the difference in the IR perturbations leading
to $D_5^{(+)}$ and $A_5^{(+)}$, the first thought that comes to mind
is to blame the unspecified `$\ldots$' for this difference.
However, this would imply that the ${\cal O}(r^{-1})$
corrections to the IR limits of {\it all}
the energy scaling functions in the two theories are the same,
whereas based on the flows described above one would expect
this only for the $\ZZ_2$-even sector!
We would like to suggest that actually the IR perturbations
are already different in the first $\phi_{3,1}$ term, the difference
being the {\it sign} of the coupling $g$.
[Note that this sign difference is also the only way to avoid the
following potential contradiction: If the leading IR perturbation
were the same, then conditions a) and b) mentioned earlier and the
equality of $E_0(R)$ in $A_5^{(+)}$ and $D_5^{(+)}$ would imply
that these are described by exactly
the same IR perturbation of $A_4$,
{\it i.e.}~are the same theories, which is wrong.
There is however one caveat, namely, if
\IRpCFT\ contains terms of the form $g^\alpha \int d^2 x \tilde{\phi}(x)$,
$\alpha >1$, which are distinct for $A_5^{(+)}$ and $D_5^{(+)}$, {\it and}
$\tilde{\phi}$ is such that it does not appear in the (repeated) fusion of
$\phi_{3,1}$ with itself. We consider the existence of such terms
unlikely, but do not really have an argument to exclude them.]

Consider, in fact, the perturbative
expansion of scaling functions based on \IRpCFT ~for $m=5$ with the
`$\ldots$' ignored, namely $A_{A_4}+g\int \phi_{3,1}$.
The resulting CPT coefficients are of the form
\anCPT\ with
$\phi_{1,3}$     replaced by $\phi_{3,1}$, $y_m$ by $2-d_{3,1}^{(4)}=-1$, and
$\kappa_m$ by $\kappa_{{\scriptscriptstyle {\rm IR}}}$.
Now as already noted in \rourv, in $A_4$
$\phi_{3,1}$ is the same as $\phi_{1,4}$, the latter being a
$\tilde{\ZZ}_2$-odd member of the $(1,q)$-operator subalgebra of the model.
(This $\tilde{\ZZ}_2$
symmetry, corresponding to the self-duality of the tricritical Ising
lattice model,
has nothing to do with the $\ZZ_2$ symmetry discussed so far in the paper,
which corresponds to ``spin reversal''. In particular, the $(1,q)$-operator
subalgebra of $A_4$,
containing both $\tilde{\ZZ}_2$-even and $\tilde{\ZZ}_2$-odd fields,
constitutes the whole $\ZZ_2$-even sector of the model.)
Therefore all correlators $\langle \phi~\prod_1^n \phi_{1,4}~\phi \rangle$
with a $\ZZ_2$-even field $\phi$ ({\it i.e.}~of definite
$\tilde{\ZZ}_2$-charge) and $n$ odd vanish in $A_4$, and so do the
corresponding CPT coefficients. As a result, the whole perturbative expansion
of scaling functions corresponding to the $\ZZ_2$-even IR fields is
independent of the sign of $g$, and is in fact in powers of $r^{-2}$
(modulo an $r^{-2} \ln r$ term, see below).
On the other hand, for the $\ZZ_2$-odd IR fields in the conformal families
of $\phi_{2,2}$ and $\phi_{2,1}$ (not belonging to the $(1,q)$ subalgebra,
{\it i.e.}~not being $\tilde{\ZZ}_2$ eigenstates)
the sign of $g$ {\it does} matter.

Of course the above arguments have to
be generalized to include additional IR perturbing fields.
[One may try to exploit our observations regarding the
finite-volume spectra of the two theories $X_5^{(+)}$ when trying to determine
these additional perturbations.
For example, allowed perturbing fields are all the
spinless descendants of $\phi_{3,1}$, whose RG eigenvalues
are necessarily odd negative integers, and all spinless descendants of the
$\tilde{\ZZ}_2$-even identity field, whose RG eigenvalues are even negative
integers.]
Still, they are very suggestive and make us believe that in the
(conjectured) RG flows from $A_5$ and $D_5$
to the common IR fixed point $A_4$, the latter
is approached from exactly opposite directions.

An important lesson of the above discussion is the following. For $m\ge 6$
the $n$-point functions of the IR-perturbing field $\phi_{3,1}$ in $X_{m-1}$
do {\it not} vanish for all odd $n$. Hence, consistency with the UV-CPT
observation that the $\ZZ_2$-even sectors of $A_m^{(+)}$ and $D_m^{(+)}$
have the same finite-volume spectrum, seems to require that
the IR perturbation of $X_{m-1}$ leading to $A_m^{(+)}$ and $D_m^{(+)}$
is the same, also in sign.
Therefore, $A_m^{(+)}: A_m \to A_{m-1}$ and
$A_m^{(+)} \neq D_m^{(+)}$ clearly imply
that $D_m$ flows to $D_{m-1}$, not $A_{m-1}$,
for $m\ge 6$.  By the same argument, the CFT
at the IR limit of $E_m^{(+)}$, $m=12,18,30$, must be $E_{m-1}$.

To conclude this section we briefly describe results of the TBA approach
relevant to the IR behaviour of $X_5^{(+)}$. In this case, the energy
scaling functions for which integral equations have been conjectured
are $e(A_5^{(+)},(1,1)|r)=e(D_5^{(+)},(1,1)|r)$ and
$e(D_5^{(+)},(3,3)^-|r)$. Solving numerically the TBA equations, we
found~\rourv ~for $r\gg 1$
\eqn\gsIRexp{ e(X_5^{(+)},(1,1)|r)~=~ -{7\over 120}-0.02723(2)~r^{-2}\ln r
      + 0.0173(2)~r^{-2} +\ldots}
for the ground state, and for the excitation describing the flow
$\phi_{3,3}^-\to\phi_{2,2}$ in $D_5^{(+)}$ we estimate
\eqn\excIRexp{ e(D_5^{(+)},(3,3)^-|r)~=~{1\over 60}+0.03333(1)~r^{-1}+\ldots~.}
These fits are based on numerical results for the scaling functions
evaluated at $200 \leq r \leq 600$.

We should emphasize that in general it is much more difficult to perform the
fits of the TBA scaling functions at large $r$, leading  in our case to
\gsIRexp--\excIRexp, than those of the small $r$ dependence leading to
UV-CPT coefficients (Table 2, for instance). We were therefore able to
determine with some confidence only the leading (asymptotic) expansion terms
indicated in \gsIRexp--\excIRexp. Still, the basic features predicted by
IR-CPT  based on \IRpCFT\ are already noticable, namely the
expansion in powers of $r^{-1}$.
The presence of the $r^{-2}\ln r$ term in \gsIRexp\ indicates the divergence
of the second CPT coefficient (multiplying $r^{-2}$) even after its UV
regularization via analytic continuation in
$y_{{\scriptscriptstyle {\rm IR}}}=2-d_{3,1}^{(4)}$.
As in UV-CPT, the same non-analytic in
$r^{y_{{\scriptscriptstyle {\rm IR}}}}$ term is expected to appear in
the large $r$ expansions of all the scaling functions in the model, in
particular in \excIRexp. Unfortunately, the accuracy of our numerical
results for $e(D_5^{(+)},(3,3)^-|r\gg 1)$ does not allow us to test this
prediction.

There is however one intriguing observation we can make based on
\gsIRexp--\excIRexp, which
can be taken (with a grain of salt) as a consistency check of these results.
 First, using \aoneCPT\ with ~$\IC_{(2,2)(3,1)(2,2)}^{(A_4)}={1\over 56}$
{}~\rDotFatiii\ we extract from the ${\cal O}(r^{-1})$ term in \excIRexp\ the
value $\kappa_{{\scriptscriptstyle {\rm IR}}}
=14\pi^{-2}\cdot 0.03333(1)=0.04728(2)$.
We now extend the observations made in the appendix regarding
logarithmic terms in UV-CPT to the case at hand, using the recipe~(A.4)
to treat the logarithmic divergence of the leading IR-CPT expansion term
of the ground state. The divergence is now due to the simple pole at $y=-1$
in eq.~(A.1), whose residue is
$R_0=-\pi^4\kappa_{{\scriptscriptstyle {\rm IR}}}^2/16$. Eq.~(A.4) then
expresses the coefficient of the $r^{-2}\ln r$ term in \gsIRexp\ as
$B_{{\scriptscriptstyle {\rm IR}}}=\alpha R_0$, and from the
numerically obtained values of
$\kappa_{{\scriptscriptstyle {\rm IR}}}$ and $B=-0.02723(2)$ we
compute $\alpha=2.001(3)$.
After reading the appendix, the reader will hopefully believe
that in fact $\alpha=2$, which leads to the (conjectured) exact relation
\eqn\IRreln{ B_{{\scriptscriptstyle {\rm IR}}} =
 - {\pi^4 \kappa_{{\scriptscriptstyle {\rm IR}}}^2 \over 8}~~.}

\newsec{Discussion}

We have presented evidence for the existence of RG flows between members
of the $D$-series of minimal unitary CFTs. These flows are induced by
perturbation of the UV CFT by the least relevant (spinless) field in the
model, a perturbation that preserves the integrability of the theory
away from criticality.
We have studied the explicit flow of scaling dimensions from the UV to
the IR CFT for various operators. For the RG flows $D_5 \to A_4$ and
$A_5 \to A_4$ we have presented strong evidence that they approach
$A_4$ from exactly opposite directions.

If there is one general lesson to be learned from this and other related
studies, we think it is this: In trying to understand integrable
QFTs interpolating between
different RG fixed points, the study of the finite-volume spectrum can
be a very powerful tool.
 The small- and large-volume behaviour of the
spectrum gives information about the UV and IR fixed points. It can
be studied with a variety of analytical and numerical techniques, like
the TBA, UV- and IR-CPT (as well as the ``truncated conformal space
approach''~\rYZ\     which
we have not utilized here).
Admittedly, ``TBA'' integral equations (in particular for excited states)
often involve some guesswork, but the structural regularities observed
in the known TBA equations  make such guesses quite natural, in many cases.
The TBA approach is
non-perturbative and the equations for excited states show quantitatively
how certain fields flow from the UV all the way to the IR; we think
this is at least as interesting as
the LG analyses   
(which in many cases are only qualitative, at best)      
and the perturbative RG
calculations which only apply if the theory is ``close'' to a model that
can be expressed in terms of (compactified, perhaps) free fields.

We would like to comment on an interesting consequence of \conj\
regarding the ``staircase model''~\rstairs\ of Al.~Zamolodchikov.
This model is given by a one-parameter ($\tht_0$)
family of diagonal $S$-matrix theories
(the status of the underlying QFTs is not clear) describing the scattering
of a single particle. The TBA analysis of the model
shows~\rstairs\ that as the parameter $\tht_0$ gets larger, the
ground state energy scaling function approximates better and better
{\it all} the functions $e(A_m^{+},(1,1)|r)$, $m=3,4,5,\ldots$,
``stringed together'' (see~\rstairs\ for the precise meaning of this
statement). Therefore in the limit $\tht_0\to\infty$ the model
describes, in some sense, the whole RG trajectory of flows between
the unitary minimal models starting at $m=\infty$
($c_{{\scriptscriptstyle {\rm UV}}}=1$) and going all the way down
to the Ising CFT $m=3$ and finally to the trivial massive IR fixed point
($c_{{\scriptscriptstyle {\rm IR}}}=0$).

In~\rFend\ the TBA equations for the lowest energy
level of the model in the sector of anti-periodic boundary conditions
were considered. Using the results of~\rourv\ the corresponding
scaling function was seen to approximate (as $\tht_0\to\infty$)
the whole set of $e(A_m^{(+)},({m\over 2},{m\over 2})|r)$ for $m$ even
and $e^{(+)}(m|r)$ for $m$ odd, stringed together (in an alternating pattern).
As was pointed out
in~\rFend, this does not approximate a sequence of excitations in
the $A$-series flow. But now having the new insight \conj, our
interpretation is that the sequence of excitations involved is
in the $D$-series, describing the flow of spin fields
{}~$\ldots \to \phi_{2\rho+1,2\rho+1}^- \to \phi_{2\rho,2\rho} \to
\phi_{2\rho,2\rho} \to \phi_{2\rho-1,2\rho-1}^- \to
\phi_{2\rho-1,2\rho-1}^- \to \ldots \to \phi_{3,3}^- \to \phi_{2,2}
\to \phi_{2,2}$~~ in ~$\ldots \to D_{4\rho+1} \to D_{4\rho} \to D_{4\rho-1}
\to D_{4\rho-2} \to D_{4\rho-3} \to \ldots \to D_5 \to A_4 \to A_3$,
respectively, and finally in the last step the relevant excitation
is $e(A_3^{(-)},(2,2)|r)$ that becomes degenerate with the ground
state $e(A_3^{(+)},(1,1)|r)=e(A_3^{(-)},(1,1)|r)$ in the IR limit $r\to\infty$.

The significance of this observation
is not yet clear. In particular, does it indicate that
the staircase model actually interpolates (asymptotically)
between all the CFTs in the $D$-series  rather than in the $A$-series
(recall that the finite-volume ground state energy is the same in
the $A$- and $D$-flows)? Or, does it indicate the
existence of an ``orbifolded staircase model'' approximating the $D$-flow,
with the modified TBA equations discussed above describing an excitation in
this model?  Obviously, to answer these questions a better
understanding  of the staircase model itself is required.

Related to this, the reader might wonder why the simplest possible
integral equation for an excited state in $X_m^{(+)}$,
namely the type $t=(-1,\ldots,-1)$
system discussed in sect.~2,  describes an excitation in
$D_m^{(+)}$ and not $A_m^{(+)}$ when
$m$ is odd (for $m$ even the
excitation energies in question coincide in the theories $A_m^{(+)}$ and
$D_m^{(+)}$). We do not have a satisfactory explanation of this at present,
it is basically an ``empirical'' observation.
However, we note that in the study~\rorbif\ of off-critical orbifolds of
the   lattice models related to $A_m^{(\pm)}$
a fundamental difference was observed between the cases
of $m$ even and $m$ odd\foot{We thank P.~Fendley for this comment.}
(specifically, applying the ``orbifolding'' procedure~\rorbif\ to the lattice
models of $A_m^{(\pm)}$ leads to
a model in the $D$-series, namely $D_{{\scsc {m+3 \over 2}}}^{(\pm)}$,
only when $m\ge 5$ is odd).
We think that this work, together with~\rFend,
might be helpful in understanding the issue.

\medskip
\bigskip
\centerline{\bf Note added}
\medskip
While  putting the final touches on this paper
we received~\rRav, which
presents a perturbative argument for flows between
minimal models high up ($m\gg 1$) in the $D$-series.
It is based on~\rGhoSen, where the difference of the torus
partition functions
of $A_m^{(+)}$ at its UV and IR fixed points is calculated
to leading order in $1/m$. The fact that flows in the $D$-series can be
distinguished from those in the $A$-series already to leading order, turns
out to follow essentially from the fact (cf.~our sect.~3) that there are
half as many spinless primary fields in $D_m$ than in $A_m$, for large $m$.
To leading order, at least, it is not possible to follow the flow of
individual operators using this method.

\bigskip
\vbox{\centerline{\bf Acknowledgements}
\medskip
We would like to thank P.~Fendley, J.~Fuchs, K.~Li, K.~Schoutens for
useful discussions. The work of T.R.K.~is supported
by NSERC and the NSF. That of E.M. by the NSF, grant no.~91-08054. }

\bigskip
\medskip
\appendix{A}{Logarithmic terms in CPT}

We present some  observations regarding
logarithmic terms in the small $r$ expansion of energy scaling functions
in certain perturbed CFTs. As mentioned in sect.~2, the appearance of
such terms is indicated within the framework of CPT by the
divergence, even after analytic continuation in $y$
(the RG eigenvalue of the perturbing field),
of certain CPT coefficients in the perturbative expansion in powers of $r^y$.
It seems that CPT is completely useless in
such cases, in particular there is no way to predict the form of the
resulting non-analyticity in $r^y$  in the
true non-perturbative answer.
However, by considering cases where
exact non-perturbative information is available, we empirically arrive at a
recipe which --- though incomplete, as we shall see --- allows one to
obtain some interesting (conjectured) results.

Consider a generic perturbed (unitary) CFT defined by an action of the
form \pCFT, with a single perturbing field $\phi$
of RG eigenvalue $0<y<2$ replacing $\phi_{1,3}$ there.
(We assume that $\phi$ is the most relevant field --- except for the
identity --- in some subalgebra of the operator algebra of the model,
so that no nontrivial fields  have to be added~\rTanig\ as counterterms to the
perturbation when renormalizing the theory.)
 For the ground state energy scaling function $e_0(r)$ one
can obtain the two leading CPT coefficients analytically from \anCPT ~(see
{\it e.g.}~\rourv ). Namely, in the CPT expansion
$e_0(r)=-{c\over 12}+\sum_{n=1}^\infty a_n r^{ny}$ we have $a_1=0$
(cf.~\aoneCPT ) and
\eqn\atwo{a_2(y) ~=~ -{\textstyle {1\over 4}} (2\pi)^{2(1-y)}~\kappa^2~
     \gamma^2(1-{\textstyle {y\over 2}})~\gamma(y-1)~~,}
\eqn\athree{a_3(y) ~=~ {\textstyle {1\over 48}} (2\pi)^{3(1-y)}~\kappa^3~
     ~\IC_{\phi\phi\phi}
  ~\gamma^3({\textstyle {2-y\over 4}})~\gamma({\textstyle {3y-2\over 4}})~~.}
Here $\gamma(s)=\Gamma(s)/\Gamma(1-s)$ and ~$\IC_{\phi\phi\phi}$ is an operator
product coefficient. (In order not to complicate the notation
we considered only the perturbation with positive $\lam$.)

Eqs.~\atwo\ and \athree\ actually give the analytic continuation in $y$ of the
corresponding integrals \anCPT, which converge only when $1<y<2$ and
${2\over 3}<y<2$, respectively.
Known non-perturbative
results of the TBA approach~\rRSOStba\rmlesstba\rourv\ lead one to
conclude that
these analytic continuations (and analogous ones for higher CPT coefficients
that we are unable to compute explicitly) provide a consistent
renormalization scheme --- not only UV regularization ---  for all
$y \in (0,2)$ different from ${2\over N}$, where $N$ is a positive integer.
(We are not aware of a rigorous proof of this fact, though it is highly
plausible, since this kind of renormalization scheme
is essentially equivalent to dimensional regularization
with minimal subtraction in ordinary QFT.)

But what if $y={2\over N}$, $N=2,3,\ldots$?
Then the analytically continued $a_\sN$ has a simple pole at ${2\over N}$
($a_n$ with $n<N$ are finite after analytic continuation and those with
$n>N$ are convergent to begin with --- see~\rouriii\rourv\ and references
therein for detailed discussion).
The simple pole indicates that if one introduces
a UV cutoff $\varepsilon$ to regularize the divergence, then the integral
for $a_\sN$ diverges logarithmically when $\varepsilon\to 0$, unlike powerlike
divergences in $\varepsilon$ that are encountered in $a_n$ with $n<N$.
There are two cases, the theories $A_3^{(\pm)}$
and $A_5^{(\pm)}$ where $y={2\over N}$ with $N=2,3$, respectively, in which
in order to obtain the known exact TBA results
the following recipe can be used:
Expand (the analytically continued) $a_\sN$ in Laurent series around
${2\over N}$
\eqn\aNLaur{ a_\sN(y)~=~
   \sum_{k=-1}^\infty R_{k+1}~(y-{\textstyle {2\over N}})^k~~;}
then in the CPT expansion of $e_0(r)$ replace
\eqn\recipe{ a_\sN r^{\sN y}~=~ a_\sN r^2
   ~~~~~~ {\rm with} ~~~~~~~
      (\alpha R_0  \ln \beta r + R_1)~r^2 }
where $\alpha,\beta$ are certain (real) constants.

The remarkable fact
is that $\alpha$ comes out to be a ``nice'' rational number --- see
below.\foot{This is also the case in
a similar recipe that was proposed by Dotsenko~\rDotsCPT\ for
treating divergences in the CPT expansion of the spin-spin correlation
function in the so-called Ising field theory (IFT) $A_3^{(\pm)}$ on the
{\it plane}. There the scaling variable analogous to our $r$ is proportional
to the separation of the fields rather than the volume of space.
A major difference between Dotsenko's problem and ours is that his recipe
is meant to cure {\it IR} divergences that are present in {\it infinitely}
many orders of CPT, whereas we deal here with {\it UV} divergences of only
{\it finitely} many terms in the perturbative expansion. This is manifested
in the non-perturbative result~\rIFT\ for the problem in~\rDotsCPT\ by
the existence of infinitely many power-log terms in the expansion, whereas
in the finite-volume energies there is (apparently) at
most a single
such term. \nl}
Moreover, the recipe is supposed to be ``universal'' in the sense that
one has to use the same $\alpha$ and $\beta$ when treating the
logarithmically divergent $N$-th CPT coefficents for {\it all} the energy
scaling functions in the given model. This requires that we keep the $R_1$
term in \recipe, which could have been absorbed in $\beta$
as long as only the ground state is concerned,
allowing for different coefficients
of $r^2$ in different scaling functions
(note, though, that the coefficient of the
$r^2 \ln r$ term is the same in all of them, as follows from the fact
that the ``strength'' of a possible UV divergence in \anCPT\ is independent
of the in- and out-fields $\phi_{p,q}$).

We first demonstrate the use of our recipe for the IFT $A_3^{(\pm)}$.
In this case the complete finite-volume spectrum is
known analytically (see sect.~6 of~\rouriii\ and references therein).  For
the ground state scaling function, in particular,
\eqn\IFTgs{e_0(r)~=~ -{1\over 24} -{\kappa^2\over 2}~ r^2 \ln r +
  {\kappa^2 \over 2} \ln\left( \pi e^{\shalf-\gamma_\sE} \right) r^2
  + {\cal O}(r^4)~,}
where $\gamma_\sE=0.577215\ldots$ is Euler's constant and
$\kappa={1\over 2\pi}$. ($\kappa$ is exactly known from the Lagrangian
formulation of
$A_3^{(+)}$ as obtained from a free massive Majorana fermion through a
``GSO projection''~\rourv, since
the perturbation $\lam \phi_{1,3}$ leading to
$A_3^{(+)}$ is equivalent there to the fermion mass term
${M\over 2\pi}i\bar{\psi}\psi$, in conventional complex notation.)
Using this exact result we conclude from \atwo\ and \recipe\ that
\eqn\IFTab{ \alpha=2~~~~~,~~~~~~\beta=\shalf e^{\gamma_\sE-\shalf} }
in IFT. We verified that our recipe gives the correct $r^2$ term
also in the scaling function corresponding to the field
 $T\bar{T}$ of the UV Ising CFT (the integration
of the relevant correlator can be performed analytically using
results of sect.~4.2 of~\rourv ).

We turn to the ground state in $A_5^{(+)}$, for which Al.~Zamolodchikov
evaluated the coefficient of the $r^2 \ln r$ term analytically from the
TBA equations he proposed~\rmlesstba. Together with our numerical
results~\rourv, the leading expansion terms obtained in the TBA approach read
(recall that $y={2\over 3}$ here)
\eqn\Avgs{e_0(r)~=~ -{1\over 15} +0.016781684(2)~r^{4/3}
  +{1\over 12\pi^2} r^2 \ln r - 0.00926790(5)~r^2 + {\cal O}(r^{8/3})~.}
Now comparing $a_2$ here with the CPT prediction \atwo\ we
obtained $\kappa_5=0.130234474(2)$ of Table~1 (actually this value is
deduced from our slightly more accurate results for $a_2$ in $A_5^{(-)}$).
Applying our recipe \recipe\ to \athree ~(in our case
{}~$\IC_{\phi\phi\phi}=~\IC_{(1,3)(1,3)(1,3)}={2\sqrt{2}\over 3}$
{}~~\rDotFatiii), we then obtain for the coefficient of the $r^2 \ln r$
term
\eqn\Avlog{ {1\over 12\pi^2} ~=~ {2\pi\sqrt{2}\over 27}~
   \gamma^{3}({\textstyle {1\over 3}})~\alpha~\kappa_5^3 ~.}
Using the numerical value for $\kappa_5$ this relation gives
$\alpha=1.49999998(6)$, which according to the
``principle of nice numbers'' is nothing but ${3\over 2}$.
The   numerical value for
$\beta$, which can be obtained using the $a_3$ that we found,
has not yet been illuminating to us.

{\it Assuming} the validity of \recipe\ with $\alpha={3\over 2}$,
we can invert the argument and obtain a new result. Namely
the exact expression for $\kappa_5$ that we
presented  (just alluding to the ``derivation'' given here) in~\rourv :
\eqn\kappav{ \kappa_5 ~=~
  {18^{1/6} \gamma({\textstyle {2\over 3}}) \over 2\pi} ~=~
    0.13023447336\ldots ~~.}
The same exact expression has been also given recently
in~\rZNtba.\foot{The $\kappa_5$ read off from eq.~(33) of~\rZNtba\
differs from ours by a factor of $\sqrt{2}$
due to a different normalization of the perturbing field. \nl}
There the authors study certain integrable perturbations
of the $Z_\sN$-parafermion CFTs, the case $N=3$ corresponding to $D_5^{(\pm)}$
where $\kappa$ is same as in $A_5^{(\pm)}$.
Though conceptually different, their analysis also resorts in the last
step to some (ad-hoc) prescription for obtaining a finite number out of
a logarithmically divergent integral --- see eq.~(32) there.

\vfill
\eject
\listrefs
\bye\end